\documentclass[pra,preprint]{revtex4}
\usepackage{graphicx}
\usepackage{verbatim}
\usepackage{amsmath}
\usepackage{amssymb}

\begin{document}

\newcommand{\bra}[1]{\langle{#1}|}
\newcommand{\ket}[1]{|{#1}\rangle}
\newcommand{\eq}[1]{(\ref{#1})}
\newcommand{\id}{\openone}

\newcommand{\wa}{\omega_{a}}
\newcommand{\twa}{\tilde\omega_{a}}
\newcommand{\wrr}{\omega_{r}}
\newcommand{\rf}{\mathrm{rf}}
\newcommand{\spec}{\mathrm{s}}
\newcommand{\w}[1]{\omega_{#1}}
\newcommand{\wdr}[1]{\omega_{d_{#1}}}
\newcommand{\ad}{a^\dag}
\newcommand{\aop}{a}
\newcommand{\veps}{\varepsilon}

\title{Superconducting qubits}

\author{Alexandre Zagoskin}
\affiliation{Department of Physics and Astronomy, University of British Columbia, Vancouver, British Columbia, V6T 1Z1 Canada}
\affiliation{Digital Materials Lab., Frontier Research System, RIKEN (The Institute of Physical and Chemical Research), Wako-shi, Saitama, 351-0198 Japan}
\affiliation{Department of Physics, Loughborough University, Loughborough, Leicestershire, LE11 3TU United Kingdom}
\author{Alexandre Blais}
\affiliation{Departement de Physique et Regroupement Quebecois sur les Materiaux de Pointe, Universite de Sherbrooke, Sherbrooke, Quebec,  J1K 2R1 Canada}

 \maketitle

\section{Introduction}

{\em General requirements.---}
From a physicist's standpoint, the most interesting part of quantum computing research may well be the possibility to probe the boundary between the quantum and the classical worlds.  The more macroscopic are the structures involved, the better.  So far, the most ``macroscopic" qubit prototypes that have been studied in the laboratory are certain kinds of superconducting qubits~\cite{You2005a,Leggett2002a,Blatter2003,Clarke2003,Mooij2005}. To get a feeling for how macroscopic these systems can be, the states of {\em flux} qubits which are brought in a quantum superposition corresponds to currents composed of as much as $10^5 - 10^6$ electrons flowing in opposite directions in a superconducting loop.  Non-superconducting qubits realized so far are all essentially {\em micro}scopic.

The advantage of quantum superconducting circuits (QSC) as implementation of qubits is, first and foremost, due to the macroscopically quantum coherent ground state of superconductors, which (a) supports non-dissipative flow of electrical current and (b) suppresses or outright eliminates low-energy elementary excitations. The latter property counterweighs the dangerous effects of a huge number degrees of freedom, which exist in the solid state and which would otherwise severely limit or totally destroy the quantum coherence necessary for the operation of a quantum computer. In this way the primary advantage of QSC allows the realization of the secondary one: as a solid state based device, a superconducting qubit and a more complex QSC can be more easily scaled up, can have significant density, and can be produced using a well developed set of design and fabrication methods.  Moreover, since these circuits are macroscopic, it can be simpler to manipulate and read their state.

The necessity to operate at low temperatures is not a disadvantage of QSC, since almost all the quantum information processing requires low temperatures in order to suppress the effects of noise.

{\em DiVincenzo criteria.---} For the following, we will formulate an abbreviated set of requirements to a QSC serving as a register of qubits, condensed from the ``DiVincenzo criteria"\cite{DiVincenzo1998}:
\begin{enumerate}
\item{ A qubit is a two-level  quantum system, which can be controlled and read out.}
\item{ A quantum computer is a set of $N$ qubits, where, in addition, certain two-qubit operations can be performed.}
\item{ The decoherence time (i.e.~the time during which the system maintains quantum coherence) must exceed the time necessary to perform a single- or two-qubit operation by a wide margin.}  
\end{enumerate}
A qubit state can thus be described by a vector in a 2-dimensional Hilbert space spanned by the eigenstates $\{|0\rangle,|1\rangle\}$ of a Pauli operator $\sigma^z$. Usually these states are chosen to coincide with distinct classical states of the system (e.g. corresponding to different electrical charge on a capacitor; different directions of current in a circuit) and which are easily distinguished in a measurement.  ``Controlled" means that an arbitrary unitary transformation $U$ can be applied on an arbritary state $\ket{\psi}$ of the system:  $|\psi\rangle \to U|\psi\rangle$. The ``readout" requirement means that projection of $|\psi\rangle$ on some directions in $\{|0\rangle,|1\rangle\}$ can be realised and that the resulting classical states can be distinguished. The state of a $N$-qubit quantum computer belongs to a $2^N$-dimensional Hilbert space, $\bigotimes_{a=1}^N \{|0\rangle_a,|1\rangle_a\}$, and the ``two-qubit operations" demand that, for at least some qubit pairs $a,b$, one could perform arbitrary unitary transformations in the corresponding Hilbert subspaces, $\{|0\rangle_a,|1\rangle_a\}\otimes\{|0\rangle_b,|1\rangle_b\}$ in order to entangle these qubits. The last requirement, that of a long enough decoherence time, is necessary in order to preserve the quantum information through-out a computation.  

The major success of quantum information theory has been to show that {\em quantum error correction}~\cite{knill:1997} is possible and that therefore the computation time is not limited by the coherence time. As long as the error probability per gate is below a certain threshold, one is able to do arbitrary long quantum computations. In the literature, the most often cited threshold is $10^{-4}$, but this number is extremely sensitive to the assumptions about the noise and about the structure of the computer.  It can be orders of magnitude smaller~\cite{Aharonov2006} or bigger~\cite{Knill2005}, depending on the various assumptions that are made.  Unfortunately, the number of $10^{-4}$ is widely cited, irrespective of these important assumptions.

The DiVincenzo requirements 1. and 2. are clearly satisfied by a system whose ``pseudospin" Hamiltonian is
\begin{equation}
H = \sum_a \vec{h}_a\cdot\vec{\sigma}_a + \sum_{a,b} \vec{\sigma}_a \cdot {\bf J}_{ab} \cdot \vec{\sigma}_b + \dots, \label{eq_pseudospin}
\end{equation}
where $\vec{\sigma}_a = (\sigma^x_a,\sigma^y_a,\sigma^z_a)$ is the vector of  Pauli operators acting on qubit $a$.  In this expression, $\vec{h}_a$ (the effective field acting on qubit $a$) and ${\bf J}_{ab}$ (the strength of qubit-qubit coupling) should be externally controllable to sufficient extent.

{\em Alternative approaches to quantum computing.---}
It is not our goal to discuss here the origin and the necessity of DiVincenzo criteria, for which one is better directed to handbooks on quantum computing (e.g. \cite{Hirvensalo1998,Bouwmeester2000}). Nevertheless it is useful to indicate certain approaches to quantum computing which allow to simplify these requirements, and would be especially well suited for an implementation using QSCs.

The standard approach to quantum computing is based on {\em quantum circuits}, meaning the consecutive application of quantum gates (one- and two-qubit unitary operations) to the system, leading in the end  to a highly entangled and fragile quantum state, where the solution to the problem is encoded. (A good illustration of the complexity of realizing a quantum circuit is given by the famous paper~\cite{Vandersypen2001}, reporting the factorization of the number 15 using the room-temperature NMR approach. The operation required the application of around 300 $\pi$-pulses and  30 $\pi/2$-pulses.) The DiVincenzo criteria reflect the contradictory requirements to a quantum computer implementing the quantum circuits approach: on the one hand, the qubits must be maximally protected from the external influences to preserve quantum coherence; on the other hand, the need to do precise time-domain manipulations on them leads to complex control-and-readout circuitry, which will introduce decoherence into the system. 

The alternative schemes of quantum computing, such as cluster state~\cite{Briegel2001}, topological~\cite{Kitaev1997,Freedman2000,DasSarma2006}, and ground state~\cite{Mizel2001} quantum computing, avoid some of these difficulties. The most straightforward is {\em adiabatic quantum computing} (AQC)~\cite{Farhi2001,Steffen2003}. In AQC, the solution is encoded in the ground state of the system  evolving   under an adiabatically slow change of a control parameter. There are constructive proofs that for any standard quantum circuit one can realize the corresponding AQC with a local Hamiltonian and with no more than polynomial overhead  \cite{Aharonov2004,Mizel2006}. 
  Remaining in the ground state to a significant extent  protects the system against relaxation and dephasing \cite{Childs2001}. The precise time-dependent control over specific qubits is no longer an issue for AQC.  Therefore a solid state-based, and in particular, QSC-based, implementation of an AQC is feasible\cite{Kaminsky2004,Grajcar2005,Izmalkov2006}.
  One should keep in mind  that an exact realization of the prescriprions of Refs.[\onlinecite{Aharonov2004,Mizel2006}] may prove difficult in practice, while a deviation from it can lead either to exponential, instead of polynomial, overhead, or to the system ending up in one of the excited states.  The question whether useful results can be still obtained in this case was discussed in \cite{Zagoskin2007}.

{\em Josephson effect.---} 
We have seen that a qubit must have two well defined levels that are used as logical states $\ket{0}$ and $\ket{1}$.  However, in practice, very few systems in nature are defined by only two levels.  To get around this, we can take two states in a {\em nonlinear} system and treat them as an effective localized spin 1/2 (qubit).  Due to nonlinearity, the transitions to other states can be made practically negligible. In electrical circuits, the natural solution is a superconducting Josephson junction~\cite{Schmidt1997,Tinkham2004}: the only known nonlinear {\em and} nondissipative electrical element.

 Josephson junctions are formed by two superconductors separated by a weak link (a tunneling barrier, constriction, bridge, etc) (see Fig.~\ref{fig_jj0}). ``Weak" means that the probability amplitude for an electron to pass through the link (e.g. the tunneling matrix element $K$ in case of a tunneling barrier) must be small enough to be considered as a perturbation.
 The coherent  nondissipative current is carried by the Bose-condensate of Cooper pairs of electrons with (in most superconductors) opposite spins. A bulk superconductor is characterized by a position-dependent complex {\em superconducting order parameter} $\Delta$ which up to a factor can be written as 
$\sqrt{n_s} \exp(i\phi)$.  In this expression, $n_s$ is the average density of electrons in the condensate, and $\phi$ is related to the velocity of supercurrent $\vec{v}_s \propto \nabla\phi$. The supercurrent density is given by $\vec{j}_s = n_s e \vec{v}_s.$ The order parameter plays the role of macroscopic wave function of the superconductor. 

Following the approach due to Feynman (see \cite{Schmidt1997}), valid in case of a low transparency barrier, we neglect the variation of the order parameter within each superconductor and describe the system  by a two-component ``wave function",
$\Psi = (\sqrt{n_{s,1}} \exp(i\phi_1),\sqrt{n_{s,2}} \exp(i\phi_2))^{T}$. Now, assume that there is a voltage diffence $V$ between the superconductors, which means that the energy difference between them is $2eV$ (since the Cooper pair has charge $2e$). Then the Schr\"{o}dinger equation for $\Psi$ can be written as
\begin{eqnarray}
i\hbar \frac{\partial}{\partial t}\left(
\begin{array}{ll}
\sqrt{n_{s,1}} e^{i\phi_1}\\
\sqrt{n_{s,2}} e^{i\phi_2}	
\end{array}
\right) = 
\left(
\begin{array}{cc}
eV & K \\
K & -eV	
\end{array}
\right) 
\left( 
\begin{array}{ll}
\sqrt{n_{s,1}} e^{i\phi_1}\\
\sqrt{n_{s,2}} e^{i\phi_2}	
\end{array}
\right)
\end{eqnarray}
After a simple calculation   we find that 
$
dn_{s,1}/dt = (2Kn_{s,1}/\hbar) \sin(\phi_2-\phi_1);\:\: d(\phi_2-\phi_1)/dt = 2eV/\hbar,
$
that is, the celebrated formulae for the DC and AC Josephson effect:
\begin{eqnarray}
I_J = I_c \sin \phi;  \:\:\:
\dot{\phi} = 2eV/\hbar. \label{eq_Jeffect}
\end{eqnarray} 
Here the phase difference $\phi \equiv \phi_2-\phi_1$. The first equation in (\ref{eq_Jeffect}) describes the non-dissipative, equilibrium, coherent flow of electric current through the barrier, determined only by the phase difference between the superconductors and the properties of the junction (all of the later being swept into the {\em critical current} $I_c$).   In some cases, other functional dependences than $\sin\phi$ occur and these have interesting qubit applications that will be briefly discussed later on.  The second equation (\ref{eq_Jeffect}) tells that if the phase difference is not constant, the current flow will be accompanied by a finite voltage drop across the junction, and vice versa.

The equilibrium current can be obtained from the appropriate thermodynamical potential of the system: $I_J = \partial E/\partial \Phi$, where $\Phi = \Phi_0\phi/2\pi$ has the dimensionality of the magnetic flux, and $\Phi_0 = h/2e$ is the superconducting flux quantum. Therefore we add to the energy of the system its {\em Josephson energy},
\begin{equation}
E_J(\phi) = \int d\Phi I_J(\phi) = -\frac{I_c\Phi_0}{2\pi} \cos\phi \equiv -E_J \cos\phi.
\end{equation}

From the AC Josephson effect and according to the relation $V = L\dot{I}$, we see that the Josephson junction can be considered as a nonlinear inductance $L_J$:
\begin{equation}
L_J = \frac{\hbar}{2eI_c \cos\phi}.
\end{equation}
The fact that this inductance can be tuned (and even change sign) by the stationary phase difference across the junction is very useful for the applications.

As an example of this quantum inductance, consider the rf-SQUID (Fig.~\ref{fig_jj}b), i.e., a superconducting loop of inductance $L$ interrupted by a single Josephson junction.  The quantization condition~\cite{Schmidt1997}, which follows from the single-valuedness of the order parameter, connects the phase drop across the junction to the total magnetic flux through the loop via
\begin{equation}
\phi+2\pi(\Phi_x - LI_J)/\Phi_0 = 2\pi n. \label{eq_quantization}
\end{equation}
Here $\phi$ is the total phase drop across all Josephson junctions in the loop (if there are more than one), $\Phi_x$ is the external flux through the loop, and $L$ is the self-inductance of the loop. The response of the current in the loop to small changes in $\Phi_x$, and therefore the effective inductance of the SQUID, depends on the value of $\Phi_x$.   

The total energy of a Josephson junction also includes its electrostatic energy:
\begin{equation}
E = Q^2/2C + E_J(\phi) = E_C (N_C-N_g)^2 + E_J(\phi). \label{eq_Coulomb}
\end{equation}
Here $Q$ is the electric charge on the junction's capacitance, $C$, and $E_C = 4e^2/2C$ the charging energy (the Coulomb energy of one Cooper pair). The charge $Q$ is expressed through $N_C$ and $N_g$.  The former is the number of Cooper pairs having tunneled through the junction while the latter  depends on the system's configuration and its electrostatic interaction with its surroundings. E.g., when there is a voltage biased electrode of voltage $V_g$ and of  capacitance $C_g$ to the junction (see Fig.~\ref{fig_jj}a), we have $N_g = C_gV_g/2e + \delta N_g$ where $\delta N_g$ is the effective charge on the junction due to any other sources (like charged impurities) in the proximity.  While $N_C$ is a discrete number, $N_g$ can take arbitrary values.

To certain extent, $N_C$ and $\phi$ can be considered as conjugate variables, similar to momentum and position~\cite{Tinkham2004}. We can therefore quantize Eq.(\ref{eq_Coulomb}) by replacing the classical variables $\phi$ and $N_C$ with operators satisfying the commutation relation 
\begin{equation}
[\hat{N}_C,\hat\phi]=-i; \:\:\: \hat{N}_C = -i\partial_{\phi};
\label{eq_commutator}
\end{equation}
we omit the hats on the operators further on. 
The situation is in reality more complex than that, because $N_C$ cannot take negative values, and $\phi$ is only defined modulo $2\pi$ (see e.g. Ref.~\cite{Zagoskin1998}, \S 1.4.3). Nevertheless for large enough systems with $N_C \gg 1$, Eq.~(\ref{eq_commutator}) holds.
Given this, the junction Hamiltonian becomes
\begin{equation}
H =  E_C (-i\partial_{\phi}-N_g)^2 - E_J \cos\phi. \label{eq_Ham}
\end{equation}
Taking $N_g=0$ and expanding the cosine around $\phi=0$, we obtain the Hamiltonian of a linear oscillator with the eigenfrequency $\omega_0$:
$-\omega_0^2\phi = \ddot{\phi} = (i\hbar)^{-2} [[\phi,H],H]$, that is, $\hbar\omega_0 =  \sqrt{2E_CE_J}.$  In practice, this {\em plasma frequency} is in the range of tens of GHz.

\section{Basic types of superconducting qubits}

Three main classes of superconducting qubits, all based on Josephson junctions, have been theoretically studied and tested experimentally.   These are phase, flux and charge qubits.  The relation between the parameters $\hbar\omega_0, E_C$ and $E_J$ and the way these qubits are biased distinguishes between the different types.

{\em Phase qubit.---} Phase qubits ~\cite{Martinis2002,Yu2002} operate in the phase regime, which is defined by $E_J \gg E_C$.  In this situation, the Josephson term dominates the Hamiltonian (\ref{eq_Ham}).  These qubits consist of a single  current-biased Josephson junction (Fig.~\ref{fig_jj}d). 
The Hamiltonian of the system can be written as
\begin{equation}
H =  -E_C \partial_{\phi}^2 - E_J \cos\phi - \frac{I_b\Phi_0}{2\pi}\phi \equiv -E_C \partial_{\phi}^2 - E_J \left(\cos\phi + \frac{I_b}{I_c}\phi\right), \label{eq_CBJJ}
\end{equation}
where the last term describes the effect of the external bias current $I_b$. This is the Hamiltonian of a quantum particle in a tilted {\em washboard potential}, see Fig.~\ref{fig_jj}d.  Since $E_J \gg E_C$, in the absence of bias, $I_b=0$, each energy well contains many almost equidistant quantized levels, and the tunneling between different minima ($\phi = 0, \pm 2\pi, \pm 4\pi, \dots$) is negligible. 
On the other hand, the Josephson junction cannot support a nondissipative current exceeding $I_c$. Indeed, 
when $I_b>I_c,$ the potential in (\ref{eq_CBJJ}) no longer has local minima. The resulting solution with nonzero $\langle\dot{\phi}\rangle$ is called {\em resistive state}: the current is accompanied by finite voltage [see Eq.~(\ref{eq_Jeffect})], and is dissipative due to the normal resistance of the junction (which we neglected so far).

In the subcritical regime (in practice, when $I_b$ is about $0.95I_c -0.98I_c$ ) there remain only a few quantized levels in every local minimum.  The tunneling out of the first two levels in a given potential through the barrier is still small enough, and these can be taken as qubit states $|0\rangle$ and $|1\rangle$~\cite{Martinis2002,Yu2002}. When $I_b \approx I_c$ the phase $\phi \approx \pi/2$; expanding the Josephson potential in (\ref{eq_CBJJ}) to the third order in $\varphi \equiv \phi-\pi/2$, we find 
\begin{equation}
U_J(\varphi) = E_J\left[ (1-I_c/I_b) \varphi - \varphi^3/6\right]. \label{eq_cubic}
\end{equation} 
The plasma frequency of the resulting cubic oscillator (i.e.~the frequency of small oscillations in the quadratic potential approximating (\ref{eq_cubic}) near its minimum)  is $\omega_p(I_b) = \omega_0 (1-(I_b/I_c)^2)^{1/4}$.  The transition frequency between the two qubit states is $\omega_{01} \approx 0.95 \omega_p$ and depends on the bias current.
The effective qubit Hamiltonian (\ref{eq_CBJJ}) becomes simply $ H = -\frac{\hbar\omega_{01}}{2} \sigma^z$,
corresponding to a spin-1/2 in a field along the $z$ direction.  The $x$- and/or $y$- components, necessary for unitary rotations of the qubit, can be implemented by adding an oscillating component to the bias current at the qubit transition frequency $\omega_{01}$: $I_b = I_{b,dc} + I_{b,ac}^c \cos\omega_{01}t + I_{b,ac}^s \sin\omega_{01}t.$    The effective Hamiltonian becomes~\cite{Martinis2003}
\begin{equation}
H = -\frac{\hbar\omega_{01}}{2} \sigma^z +  \sqrt{ \frac{\hbar}{2\omega_{01} C}} \frac{I_{b,ac}^s}{2} \sigma^x
+  \sqrt{ \frac{\hbar}{2\omega_{01} C}} \frac{I_{b,ac}^c}{2} \sigma^y.
\end{equation}
The DC component of the bias current sets $\omega_{01}$, while the cosine and sine quadratures produce the rotations around the $x$ and the $y$ axis, respectively. The nonlinearity of the potential (\ref{eq_cubic}) is crucial, otherwise the AC current pulses would also excite transitions between higher levels, and out of the Hilbert space spanned by $|0\rangle$ and $|1\rangle$. In other words, the system would no longer be a qubit.

Advantages of phase qubits include the simplicity of design, good control over the parameters in the Hamiltonian, simple readout, weak sensitivity to charge and flux noise in the system, and scalability. One of the relative disadvantages is the sensitivity to low-frequency (i.e.~$1/f$) noise of critical and bias current.  This noise can be produced, for example, by microscopic defects in the insulators of tunneling barriers. Another is the absence of the optimal point, i.e., the degeneracy point where symmetry would protect the qubit from certain kinds of noise and, as we will see for other types of qubits, significantly increase its coherence time.  This is what has so far limited the system to short coherence times ($T_2 \sim 80$ ns~\cite{Steffen2006}).

In phase qubits $\sigma^{x,y}$-operations  must be realized with AC pulses.  This has the disadvantage of requiring complex high-frequency low-noise circuitry down to the very low operating temperatures ($\sim$ 10 mK).  On the other hand, using AC  rather than DC pulses allows for protection by low-pass filtration of the qubit from the low frequency ($1/f$) noise coming from the bias lines that can be significant.

{\em Flux qubit.---} Another possibility to realize a qubit in the phase limit $E_J \gg E_C$ is by using a degeneracy between two current-carrying states of an RF-SQUID (see Fig.~\ref{fig_jj}b)~\cite{Friedman2000}.  In writing the Hamiltonian of the system, the only difference with respect to the Hamiltonian of Eq.~\eq{eq_Ham} is that we now must take into account of the magnetic energy, but can neglect the effect of off-set charges $N_g$:
\begin{equation}
H =  -E_C \partial_{\phi}^2 - E_J \cos\phi + E_L(\phi-\phi_x)^2/2. \label{eq_RFSQUID}
\end{equation}
Here $\phi_x = 2\pi \Phi_x/\Phi_0$ is the reduced  flux through the loop, and the inductive energy scale is given by $E_L = \Phi_0^2/4\pi L$. The situation differs from the previous case because now, if $\phi_x \approx \pi$, the potential energy formed by the last two terms of Eq.~\eq{eq_RFSQUID} has two almost degenerate minima.  As illustrated in Figure~\ref{fig_jj}b, these states correspond to persistent current in the loop circulating in opposite directions, and are conveniently used as the $|0\rangle$ and $|1\rangle$ states of a qubit.

Tunneling between the two potential wells is enabled by the charge term in Eq.~\eq{eq_RFSQUID}.     As a result, in the subspace $\{|0\rangle, |1\rangle\}$  the effective qubit Hamiltonian is
\begin{equation}
H = -\frac{\epsilon}{2} \sigma^z + \frac{\Delta}{2} \sigma^x, \label{eq_fluxH}
\end{equation}
where $\epsilon \propto \langle 1|H|1\rangle - \langle 0|H|0\rangle$ is the energy bias between the two states, which is tuned by the external flux; $\Delta \propto \langle 0|H|1\rangle$ is the tunneling amplitude.  When $\epsilon \gg \Delta$, the eigenstates of (\ref{eq_fluxH}) coincide with $|0\rangle$ and $|1\rangle$. At the   degeneracy point, $\epsilon = 0$ ( $\phi_x=\pi$ ), these eigenstates are ``bonding/antibonding" states $\ket{\pm}=(|0\rangle\pm|1\rangle)/\sqrt{2}$.
The expectation value of the current circulating in the SQUID loop, $\hat{I}$, is the same in $\ket{+}$ and $\ket{-}$: $\bra{+}\hat I \ket{+}=\bra{-}\hat I \ket{-}$.  As a result, these two states cannot be distinguished by observing the current, or the magnetic field it produces.  Therefore no external degrees of freedom that couple to magnetic field can interfere with the qubit when it is operated at this  optimal point; the coherence time of the qubit can thus be significantly enhanced.  Therefore it is convenient to always work at this special operating point and use the corresponding eigenstates as effective logical states of the qubit: $\ket{0_L} = \ket{+}$, $\ket{1_L} = \ket{-}$.  While the Hamiltonian (\ref{eq_fluxH}) enables all qubit rotations with DC pulses only, transitions between these states can also be realized by applying RF flux pulses at the transition frequency, same as in a phase qubit.

Note that no difference in observed current means that it is impossible to readout the state of the qubit by looking at current or magnetic field produced by the loop. One can either move away from the optimal point for the readout, as was done in earlier experiments, or to use {\em dispersive readout} discussed later.  

The major problem with the RF-SQUID design is that in order to have a potential energy landscape with two well-defined minima, both the inductance of the SQUID loop and the Josephson energy must be large. First, this suppresses the tunneling; in 
  the experiment~\cite{Friedman2000} it was impossible to produce coherent superpositions between the lowest states in the wells, and instead the higher states, close to the top of the barrier, had to be used; transitions to these states were induced by resonant RF pulses.  Second,  due to large inductance the states $\ket{0}$ and $\ket{1}$ produce quite large magnetic fields (these two states differ by about one flux quantum $\Phi_0$) and the minute deviations from the optimal point lead to strong coupling to the environment and fast dephasing. This problem is solved in the flux qubit design  of Ref.~\cite{Orlando1999,Wal2000a},  Fig.~\ref{fig_jj}c, where the RF SQUID is replaced by a loop of negligible inductance $L$ but with {\em three} Josephson junctions.  In this case the total flux through the loop practically coincides with the external flux, and the  flux quantization condition (\ref{eq_quantization}) becomes $
 \phi_1 + \phi_2 + \phi_3 + 2\pi\Phi_x/\Phi_0 = 2\pi n 
 $
(here the phase drop across junction $i$ is $\phi_i$). It leaves only two independent phases (e.g.,  $\phi_1$ and $\phi_2$).  The Josephson energy of the system provides a 2D potential landscape,
$E_J(\phi_1,\phi_2;\phi_x) = -E_{J1} \cos\phi_1 - E_{J2}\cos\phi_2 - E_{J3}\cos(\phi_x+\phi_1+\phi_2)$, plotted in Fig.~\ref{fig_jj}c and containing suitable double well structures.  The two qubit states again produce circulating currents in the qubit loop.  However, the generated flux is now much smaller, and the two states are separated by only $\sim 10^{-3} \Phi_0$.  As a result, unwanted coupling to the environment is much weaker, with appropriately longer coherence times.  The tunneling barrier can be also lowered, enabling tunneling between the lowest states in each minimum. The Hamiltonian of this qubit can be also cast in  the form  (\ref{eq_fluxH}),  and the same optimal point considerations apply.  Note that a consistent derivation of the limit  $L\to 0$ from Eq.(\ref{eq_RFSQUID}) is rather involved~\cite{MaassenvandenBrink2005}.

While the fluxes produced by the persistent currents in $|0\rangle$ and $|1\rangle$ only differ by about $10^{-3}\Phi_0$, the difference of currents themselves still involve $\sim 10^5-10^6$ single-electron states.  A state of the form $(|0\rangle\pm|1\rangle)/\sqrt{2}$ thus involves a superposition of millions of charges flowing in two directions at once.  Superconducting flux qubits are truly macroscopic quantum systems. (A more detailed analysis of how to define what is a macroscopic quantum system is given in \cite{Leggett1980,Leggett2002}.)

Advantages of flux qubits are their weak sensitivity to charge noise, comparatively easy readout, the ability to work at or near the degeneracy point (optimal point).  Moreover, in principle,  all operations on the qubit can be performed using DC pulses only. Their main disadvantage is a very strong dependence on the junction parameters. The tunneling splitting $\Delta$ is proportional to $\sqrt{E_JE_C} \exp[-\sqrt{E_J/E_C}]$, while $E_J$ itself depends exponentially on the barrier thickness. Such fluctuations as $1/f$ noise in the critical current will thus have a big effect on the qubit.

{\em Charge qubit.---} The charge qubit (Fig.~\ref{fig_jj}a) operates in the opposite regime from the previous two types of qubits, the charge limit $E_C \gg E_J$. Now the states with definite charge, differing by a single Cooper pair, are the working states of the qubit, and the Josephson term is the perturbation. Historically, this the first superconducting qubit to have been experimentally implemented~\cite{bouchiat:98,Nakamura1999}.  This is  not completely surprising given that, in a way, it is the ``most microscopic" of them all.

In order for a single Cooper pair to make a difference, two conditions must be met~\cite{Tinkham2004}.  The charging energy $E_C \sim e^2/C$ must far exceed both the thermal energy, $k_BT$, and the linewidth of the charge states, $\Delta E \sim \hbar/RC$, due to their finite lifetime determined by the effective resistance and capacitance of the system. Otherwise, the charge states are smeared out and cannot form a good basis for a qubit.  The second condition leads to the requirement $R > \hbar/e^2 \sim 6 \:\:{\rm k}\Omega$.   As noted in Ref.~\onlinecite{Tinkham2004}, \S 7.1, this becomes a problem, because at GHz frequencies dictated by small capacitances $C$, the tunneling resistance will be effectively shunted by  small impedance $\sim 50-100 \:\:\Omega$ of the connecting leads (that is, bulk superconductors forming the banks of the Josephson junction). The charge states will be washed out by the quantum particle number fluctuations.  As a result, it is easier to observe charge effects in a small superconducting {\em island} separated by high-impedance tunnel barrier from the bulk superconductor, as shown in Fig.~\ref{fig_jj}a. 
The Hamiltonian for a charge qubit can be written as the sum of the electrostatic energy and of the Josephson energy:
\begin{equation}
H =  E_C (N_C-N_g)^2 \label{eq_Ham_C} - E_J\cos\phi.
\end{equation}
In the absence of Josephson energy, the states $|N_C = 0\rangle$ and $|N_C = 1\rangle$ are degenerate when $N_g = C_gV_g/2e=1/2$. Higher-energy states can be neglected; then on the Hilbert subspace spanned by these two states, the operator $N_C$ takes the form
$N_C =
 \frac{1}{2}(1 - \sigma^z).$
 The Josephson term  lifts the charge degeneracy; in this basis it can be written as $E_J\sigma^x/2$.
 The Hamiltonian again takes the pseudospin form
\begin{equation}
H =  -\frac{1}{2}E_C(1-2N_g)\sigma^z - \frac{E_J}{2}\sigma^x, \label{eq_Ham_charge_spin}
\end{equation}
allowing all unitary rotations with DC pulses only. At the charge degeneracy point, $N_g=1/2$, the eigenstates of this Hamiltonian are coherent superpositions of states $(|0\rangle\pm|1\rangle)/\sqrt{2}$ differing by a single Cooper pair.  Like in the flux case, it is best to operate at this optimal point, and it is convenient to make a transformation to the basis of logical qubit states $(|0\rangle\pm|1\rangle)/\sqrt{2}$:    $\sigma^x \rightarrow -\sigma^z$ and $\sigma^z \rightarrow \sigma^x$ .  Then, denoting $N_g=1/2 + N_{g,ac} \cos(\omega t)$, we have
\begin{equation}
H =  \frac{E_J}{2}\sigma^z + Ec N_{g,ac} \cos(\omega t) \sigma^x.
\label{eq_Ham_charge_spin_sweet}
\end{equation}
Now the Josephson energy plays the role of transition frequency for the qubit and AC voltage $N_{g,ac}$ can be used to induce transitions between these states.  

The advantages of the charge qubit operating at the optimal point are similar to those of the flux qubit.  The former however does not have the exponential dependance on parameters that the flux qubit has.  One of its main disadvantages is very strong sensitivity to charge noise.  This can be mitigated to some extent by working in the intermediate regime, $E_J \lesssim  E_C$, which was realized in the so-called ``quantronium"~\cite{Vion2002} and, more recently, the ``transmon" qubit~\cite{Schuster2006}.

\section{Coupling qubits}

{\em Coupling by passive elements.---} By passive elements, we refer to the circuit elements with resonance frequency much higher than the $\ket{0}$ to $\ket{1}$ transition frequency of the qubits they are coupling.  As a result, the coupling element will remain in its ground state at all times.  In this category, the most commonly studied (and experimentally exploited) coupling elements are the usual linear elements of electrical circuits: capacitances and inductances.  Simplest passive couplings for phase (a), charge (b) and flux (c) qubits are shown in Fig.~\ref{fig_coupled_passive}. 
For phase qubits, joining one electrode of each junction by a capacitance $C_x$ yields an interaction term \cite{Blais2003}
\begin{equation}
H_\mathrm{int} = \frac{\hbar\sqrt{\omega_{01,a}\omega_{01,b}} C_x}{C} \sigma^{y}_a\sigma^{y}_b,
\label{eq_Hint_phase}
\end{equation}
between the qubits $a$ and $b$.  Here $\omega_{01,j}$ is the transition frequency of qubit $j$.  If the difference between the qubit eigenfrequencies, $\omega_{01,a}-\omega_{01,b}$, is much bigger than the coupling strength $\hbar\sqrt{\omega_{01,a}\omega_{01,b}} C_x/C$, then this interaction term acts as a perturbation on the qubits and can, for all practical purposes, be ignored.  Indeed, by moving to the interaction representation with respect to the Hamiltonian of noninteracting qubits, we will see that this term is fast oscillating and averages to zero, and is therefore negligible (so called Rotating Wave Approximation, RWA).  On the other hand, if $\omega_{01,a}=\omega_{01,b}$, the   term (\ref{eq_Hint_phase}) cannot be neglected and leads to the qubit-qubit coupling.  Therefore the coupling can be turned on and off by taking the qubits in and out of resonance with each other (for phase qubits,  by tuning the qubit bias currents).  If we use only DC controls, there is an important difference between  the  diagonal couplings  ($\sigma^{z}_a\sigma^{z}_b$) and off-diagonal ones ($\sigma^{x}_a\sigma^{x}_b$ and $\sigma^{y}_a\sigma^{y}_b$).  While the latter can be turned on and off by tuning the qubits, the former is always on.

In practice, one cannot neglect finite inductances of connecting elements (roughly  $L \sim 1$ nH per 1 $\mu$m). Therefore even in Fig.~\ref{fig_coupled_passive}a the qubits are coupled through a harmonic oscillator of frequency $\omega_\mathrm{osc} = \sqrt{1/LC_x}$. Taking the capacitance to be 1 fF~\cite{Steffen2006} and $L=$1 nH, its frequency is $\sim100$ GHz, about 10-20 times larger than the typical qubit transition frequencies. Therefore we have indeed a passive coupling and can safely neglect the finite inductance. In one experiment on capacitively coupled phase qubits with large  $C_x \sim 5$ pF the effects of the excited LC mode were  observed~\cite{Xu2005}.   

The capacitive coupling of two charge qubits, Fig.~\ref{fig_coupled_passive}b, leads to a similar coupling term
\begin{equation}
H_\mathrm{int} = \frac{e^2}{2 C_x} \sigma^{x}_a\sigma^{x}_b,
\end{equation}
to which the same considerations apply. 

Finally, Figure~\ref{fig_coupled_passive}c shows two flux qubits  coupled through a mutual inductance $M$. Since external flux leads to a change in energy bias $\epsilon \sigma^z/2$ in the single qubit Hamiltonian Eq.~\eq{eq_fluxH}, this coupling is 
\begin{equation}
H_\mathrm{int} =  2 M I_a I_b \sigma^{z}_a\sigma^{z}_b,
\end{equation}
where $I_j$ is the circulating current in qubit $j$. In the eigenbasis of the qubits at the optimal point, this also takes the form $\sigma^{x}_a\sigma^{x}_b$.

One can use a bona fide LC circuit as a passive coupler \cite{Makhlin1999,Makhlin2001,Zagoskin2006} (like in Fig.\ref{fig_coupled_active}b); now several qubits can be connected to a single circuit ({\em quantum bus}), and only those in resonance with each other will couple. 
If the qubits themselves {\em cannot} be tuned, one needs more complex passive coupling elements (e.g. \cite{Plourde2004,Maassen2005}). For flux qubits, coupling with tunable sign was recently realized  experimentally~\cite{Ploeg2006,Hime2006}. The coupling scheme used in \cite{Ploeg2006} is shown in Fig.~\ref{fig_coupled_active}c. The external flux through the coupler loop, $\Phi_C = f_C\Phi_0$, changes the phase shifts across the junctions shared between the coupler and the qubits $a$ and $b$. The resulting shift in the Josephson energy of the whole system leads to the (diagonal) qubit-qubit coupling $J \sigma^z_a\sigma^z_b$ with 
\begin{equation}
J (f_C) = \frac{\hbar}{2e} \frac{I'(f_C)}{I_c^2 - I(f_C)^2} \:I_a\:I_b.
\end{equation}
Here $I_c$ is the critical current of the shared junctions, and $I(f_C),\:I'(f_C)$ are the circulating current in the coupler loop and its derivative with respect to the coupler flux resp.; $I_{a,b}$ are the persistent currents in the qubits. The coupling sign changes with the isgn of the derivative $I'(f_C)$, and the experimental dots in Fig.~\ref{fig_coupled_active}c are in  a good agreement with the formula.

Use of AC controls opens still more possibilities.  The diagonal coupling can be effectively cancelled by the techniques of refocusing borrowed from NMR~\cite{vandersypen:2004}.  The off-diagonal coupling can be tuned with RF pulses while staying at the optimal point.  The general idea is to use the frequency control. For example, detune the qubits from each other, so that $\delta = \omega_{01,a}-\omega_{01,b}$ exceeds the coupling strength; the coupling is effectively off.  RF irradiation of each qubits at the difference frequency $\delta$ will effectively turn on the interaction~\cite{rigetti:2005}.   Several different frequency-control schemes have been recently proposed ~\cite{niskanen:2006,bertet:2006,liu:2006,liu:2006a,grajcar:2006,ashhab:2006}.

{\em Coupling by active elements.---} By these we understand the circuit elements whose typical transition frequency are comparable to those of the qubits they couple, and which therefore undergo transitions to excited states. In other words, now the coupling between the qubits is realized through their exchange of {\em real}, rather than {\em virtual}, excitations with the coupler. Examples of couplers which can be used in the active regime are shown in Figure~\ref{fig_coupled_active}.  
In Fig.~\ref{fig_coupled_active}a, charge qubits are coupled through a current-biased Josephson junction (CBJJ) ~\cite{Blais2003,plastina:2003}. Like in the case of a phase qubit, the junction can be considered a tunable  {\em nonlinear} oscillator, with only the two lowest states involved, and is coupled to either qubit through $g_j\sigma^x_{a,b}\sigma^y_{\rm osc}$.  The coupling strength $g_j$ depends on the coupling capacitance $C_{xj}$ between the qubit and the oscillator.  Since this interaction is off-diagonal, when the oscillator is strongly detuned from the qubits, the oscilator-qubit interaction is turned off.  By sequentially tuning the oscillator in and out of resonance with the two qubits, it is possible to entangle these qubits while leaving the oscillator unentangled~\cite{Blais2003}.  In this way, several qubits can be coupled to the same oscillator, which plays the role of the data bus. Since it is the coupling bus and not the qubits that are tuned, the qubits can remain at their optimal point throughout the manipulations.  
The same approach works for other types of qubits as well (Fig.~\ref{fig_coupled_active}b).

For a more exhaustive review of coupling schemes the reader is directed to Ref.~\cite{Wendin2006}.

{\em Qubits coupled to oscillators: circuit QED.---} The natural way superconducting qubits and oscillators couple to each other led to the insight that superconducting quantum circuits allow to realize an analogy to quantum optics in general~\cite{walls-milburn}, and  to strong coupling cavity quantum electrodynamics (cavity QED)~\cite{raimond:2001,mabuchi:2002} in particular.
This was pointed out by several authors~\cite{buisson:2001,marquardt:2001,Blais2003,you:2003a,Zagoskin2004}.  In this section, we will focus on the work of Ref.~\cite{blais:2004}.  As shown in Fig.~\ref{fig_circuit_QED}, this proposal is based on a charge qubit (green) fabricated inside a high quality superconducting transmission line resonator (blue). In the lumped element description, the transmission line resonator can be seen as a series of LC circuits connected to external leads by capacitances $C_0$.   The finite length $l$ of the resonator sets up its characteristic frequency. The 
electromagnetic field transmission between the input and output ports of the resonator is maximal at this frequency and its harmonics;  with $l=25$ mm, these resonances are in the microwave range.

Close to one of these resonances, the  resonator Hamiltonian is $H_{r} = \hbar\wrr \ad \aop,$
where $\wrr = 1/\sqrt{LC}$, and $a\: (a^{\dag})$ are the annihilation (creation) operators.
The qubit is biased not only by a DC or AC current (that can be applied on the input port of the resonator), but also biased by the voltage across the LC circuit.  This voltage can be written as $V_\mathrm{LC} = V^0_\mathrm{rms}(\ad+ \aop)$, where $V^0_\mathrm{rms} = \sqrt{\hbar\wrr/2C}$ is the rms value of the voltage in the ground state.   In practice, this rms voltage can be as large as a few $\mu$V~\cite{Wallraff2004,Schuster2005,Wallraff2005,Schuster2006}.  Inserting this in the Hamiltonian  of a charge qubit DC-biased at the optimal point~\eq{eq_Ham_charge_spin_sweet}, we find $H = \frac{E_j}{2}\sigma^z - g(\ad+ \aop)\sigma^x,$
where $g=eV^0_\mathrm{rms}C_g/C_\Sigma$ is the qubit-oscillator coupling strength, with $C_g$ the oscillator-qubit coupling capacitance and $C_\Sigma$ total capacitance of the qubit.  Writing $\sigma^x = \sigma^+ + \sigma^-$ and neglecting fast oscillating terms (RWA), we can write the qubit+oscillator Hamiltonian as
\begin{equation}
H = \hbar\wrr \ad \aop + \hbar\frac{\wa}{2}\sigma^z - \hbar g(\ad  \sigma^- + \sigma^+\aop),
\label{eq_Hjc}
\end{equation}
with $\hbar\wa=E_j$.  This expression is the Jaynes-Cummings Hamiltonian of quantum optics~\cite{walls-milburn} which is traditionally used to describe the physis of a two-level atom of frequency $\omega_a$ coupled to the photon field inside a cavity of frequency $\wrr$~\cite{raimond:2001,mabuchi:2002}.  Due to this correspondence one calls this architecture {\em circuit QED}.

Circuit QED has several advantages over cavity QED. First, due to the small volume of the transmission line resonator and to the large effective dipole moment of the charge qubit, the coupling strength $g$ can be made much bigger. Second, the positions of real atoms inside a 3D cavity fluctuate in space, leading to a fluctuations in the coupling $g$, which interfere with quantum information processing.  In circuit QED, the position of the qubit in the resonator is fixed. Third, it is possible to fabricate several qubits inside a single resonator and to entangle these qubits by using the resonator as a quantum bus~\cite{blais:2006}.  Circuit QED is thus interesting both for quantum information processing applications and as a tool to study new regimes of quantum optics.

\section{Readout of superconducting qubits}

To be useful as qubits, the quantum states of QSCs must allow an efficient readout.  In practice, the readout circuitry will be microfabricated in a  proximity to the qubit.  The challenge is thus to have this circuitry strongly coupled to the qubit during measurements (to have a fast measurement on the scale of the qubit's $T_1$) but strongly decoupled during computations (to prevent additional decoherence and relaxation of the qubit).

For phase qubits, the readout circuitry is an integral part of the qubit design.  The measurement is based on the large difference between the tunneling rate $\Gamma_i$ of the logical states $\ket{i}$~\cite{Martinis2002}.   As illustrated in Fig.~\ref{fig_jj}d,  to realize a measurement, the DC bias current is rapidly increased in such a way that while levels $\ket{0}$ and $\ket{1}$ remain in the well, the level $\ket{1}$ becomes very close to the top of the potential barrier, so that $\Gamma_1\gg\Gamma_0$.  As a result, if the qubit is in the state $\ket{1}$, it will rapidly decay in the continuum~\cite{cooper:2004}, producing the resistive state and thus a voltage of the order of the superconducting gap  ($\sim$ 1 mV) across the junction.  Detection of this voltage measures the qubit state. An important drawback of this is the generation of quasiparticles due to the finite voltage.  These will affect the qubit itself and lower the coherence time of neighboring qubits.  Therefore, in more recent experiments the junction forming the qubit was embedded in a loop of finite inductance. Now the system tunnels from the local potential minimum not into the continuum of states, but in a large but finite potential well.  As a result, there is no voltage drop across the junction, but  a change of flux in the qubit loop by $\Phi_0$.  The latter is detected by a DC-SQUID~\cite{Simmonds2004}. Remarkably, if the qubit is not kept biased long enough for the state $\ket{1}$ to decay with certainty, a {\em partial measurement} is realized~\cite{Katz2006}, accompanied by the probability amplitude concentrating in the state $\ket{0}$.

For charge and flux qubits, the first readout methods to be experimentally realized detected directly the charge (flux) associated with the qubit states. For charge qubits ~\cite{Nakamura1999}, this was first realized by capacitively coupling a superconducting single-electron transistor (SET)~\cite{bouchiat:98}, a highly sensitive charge meter, to the island of the qubit.  The radio-frequency version of the superconducting SET (RF-SET)~\cite{schoelkopf:1998} has also been used in more recent experiments~\cite{lehnert:2003,Duty2004,guillaume:2004}.  By working at RF rather than at DC, the RF-SET can be much faster than regular SETs and is also less sensitive to $1/f$ noise due to charge fluctuations.   
For flux qubits, the direct readout was realized by inductively coupling a DC-SQUID to the qubit~\cite{Wal2000a,Friedman2000}.  

A problem with the above mentioned approaches is that they do not work at the optimal point, where, as discussed earlier, the observable values of currents (fluxes) or charges in both qubit states coincide.  Away from the optimal point the system is exposed to extra decoherence and relaxation, especially if the qubit's environment is highly structured~\cite{Simmonds2004}. A partial solution is provided by the phase-charge readout developed by the Saclay group~\cite{Vion2002}.  
A different approach, {\em dispersive readout}, is based on the observation that, while the currents (charges) at the optimal point are the same in both qubit states, their {\em derivatives} with respect to the control parameters are not. This can be done by coupling the qubit to a resonator and measuring the latter's {\em susceptibility}, dependent on the quantum state of a qubit. In case of a relatively low-Q  lumped LC circuit far away from resonance (strong detuning) this approach was developed and implemented for flux qubits by the Jena-Vancouver collaboration~\cite{Ilichev2003,Izmalkov2004a,Grajcar2006,Ploeg2006} (with direct inductive coupling) and by the Delft group~\cite{Lupascu2004,Lupascu2005,Lupascu2006} (with the coupling through a DC SQUID). In a similar way, measurement of the quantum capacitance was also realized with lumped resonators in the very strong detuning regime with charge qubits~\cite{Sillanpaa2005,Duty2005}.

In the remainder of this section, we will discuss in more detail the dispersive readout in circuit QED, where this approach is especially effective
~\cite{Wallraff2004,Schuster2005,Wallraff2005,Schuster2006}. It works by going to the {\em dispersive regime}, where the qubit-resonator detuning $\Delta \equiv \hbar(\wa - \wrr) = 0$ is much larger than the coupling strength $g$.  In this situation, the Jaynes-Cummings Hamiltonian of Eq.~\eq{eq_Hjc} can be replaced, at the optimal point ($N_g=1/2$) and to second order in the small parameter $g/\Delta$, by the effective Hamiltonian
\begin{equation}
H_\mathrm{eff} = \hbar\left( \wrr + \chi \sigma^z \right) \ad \aop + \hbar\frac{\twa}{2} \sigma^z,
\label{eq_Hdiss}
\end{equation}
where $\twa = \wa+\chi$ and $\hbar\chi = g^2/\Delta$. The term $\chi \sigma^z \ad \aop$ in (\ref{eq_Hdiss}) acts as a shift of the cavity frequency that depends on the state of the qubit.  Indeed, in this situation the effective resonator frequency is no longer $\wrr$, but is now $\wrr\pm\chi$ depending on $\langle\sigma^z\rangle=\pm1$.  Using this frequency pull, it is possible to read out the state by measuring the phase and amplitude of an RF signal transmitted between the ports of the resonator. These quantities indeed contain information about the state of the qubit~\cite{blais:2004,Gambetta2006}.  

Dispersive readout has several important advantages over the approaches that directly measure charge or flux.  First, the qubit can be operated {\em and} measured at the optimal point. This is because it measures not the charge on the qubit's island but rather the resonator frequency pull which is related to the {\em quantum capacitance} of the qubit~\cite{blais:2004}.  Second,  it is actually at the optimal point that this quantum capacitance (or quantum inductance in the flux case) is maximally different for the two states of the qubit.  Thus, with this dispersive measurement, there is maximal signal at the optimal point.  Third, there is no energy exchange between the qubit and the resonator systems during readout, and the resonator is measured in its {\em eigenstate}.  This leads to what is known as {\em quantum non-demolition} (QND) measurement~\cite{walls-milburn} and is ideally suited for quantum information processing.  Last, but not least, the readout is only active when signal is sent at the input port of the resonator, which realises the desired coupling/decoupling requirement for a quantum readout.

\section{Further reading}

The short format of this review allows us neither to include all the important topics concerning superconducting quantum circuits, nor to cover the ones included in due detail. We try to remedy this by providing here a list of additional sources, which should be accessible based on what have been discussed here so far. 


{\em Reviews on SQC.---} A concise and accessible description of different types of superconducting qubits is provided in Ref.\onlinecite{You2005a} and, with a varied degree of generality, in \cite{Leggett2002a,Blatter2003,Clarke2003,Mooij2005}. The first comprehensive review is Ref.\onlinecite{Makhlin2001}. Recent  pedagogical reviews are Refs.~\onlinecite{Devoret2004,Wendin2006}.  The paper ~\cite{Devoret2004} describes all qubits as examples of a nonlinear quantum circuit, with an  emphasis on  the phase qubit. The review \cite{Wendin2006} contains also a description of abstract qubit manipulations (Sect.IV) and of the theoretical formalism necessary to analyze classical and quantum superconducting circuits (Sect. V and VI). Recent advances are reviewed in \cite{Geller2007,Wilhelm2007}.

{\em Superconductivity.---} Introductory information on theory of superconductivity and Josephson effect can be found in \cite{Schmidt1997}. The classic book \cite{Tinkham2004}    devotes a whole chapter (Ch.7) to the effects in small Josephson junctions  highly relevant for the qubit theory. This topic is also covered in \cite{Zagoskin1998}, Ch.4.

{\em Theoretical formalism.--} Detailed  explanation of the formalism allowing a uniform description of arbitrary quantum superconducting circuits, based on a Lagrangian approach, is given in~\cite{Devoret1997,Burkard2004}. The formalism is linked to the graph description of complex circuits, standard in electrical engineering.  Effective Hamiltonians for qubit manipulations and qubit-qubit coupling are derived in~\cite{Blatter2001,MaassenvandenBrink2005}.

{\em Decoherence and dissipation.---} Sources of decoherence in various superconducting qubits are discussed in  \cite{Han2001,Tian2002,Zagoskin2003,Wilhelm2003,Martinis2003,Wal2003,Storcz2003a,Simmonds2004,Amin2004,Duty2004a,Astafiev2004a,Martinis2005,Burkard2005,Burkard2005a,Astafiev2006}.

{\em Recent experimental progress.---} After significant improvements to the initial design, two-qubit quantum operations (equivalent to a universal $\sqrt{\rm iSWAP}$ gate) were demonstrated in two capacitively coupled {\em phase qubits}~\cite{McDermott2005,Steffen2006}.  In particular, in Ref.~\onlinecite{Steffen2006}, quantum state tomography~\cite{Nielsen2000} was realized to extract the density matrix representing the state of the coupled qubits.  An  experiment with  two phase qubits capacitively coupled through a series $LC$ resonator showed the evidence for entanglement between the qubits and the resonator mode~\cite{Xu2005}.  In single phase qubits the behaviour of quantum two-level systems (TLS), which are believed to be the main source of $1/f$ noise, was investigated, and coherent beats between a TLS and a qubit were observed~\cite{Simmonds2004,Martinis2005}. This led to a proposal~\cite{Zagoskin2006} to use such TLSs as naturally formed qubits, with the phase qubit playing the role of a quantum bus.

With {\em flux qubits}, much effort was devoted to the development of structures suitable for adiabatic quantum computing. After the initial proposals \cite{Kaminsky2004,Grajcar2005}, coupling between two~\cite{Grajcar2005c}, three~\cite{Izmalkov2006} and four~\cite{Grajcar2006} flux qubits was realized in the quantum regime, and entangled eigenstates were formed. We already mentioned tunable coupling between two flux qubits~\cite{Ploeg2006,Hime2006,Harris2007}. {\em Time- domain} tunable coupling between flux qubits and a simple quantum protocol were experimentally realized in Ref.\cite{Niskanen2007}.
{\em Multiphoton} Rabi oscillations were observed in  a flux qubit~\cite{Saito2005}. 
In Ref.~\cite{Petrashov2005} a flux qubit was measured using the {\em Andreev probe}, that is, measuring the   electrical conductance in a non-superconducting (in this particular case, silver) wire connected by superconducting leads to two points of the qubit loop. The normal current through the wire is carried by so-called Andreev levels in a normal conductor sandwiched between superconductors (see ~\cite{Zagoskin1998}, \S A.2). The conductance is therefore sensitive to the phase difference, and allows to determine the state of the qubit with a minimal backaction.

The longest relaxation and dephasing times for  charge qubits  (8 $\mu$s and 500 ns, respectively) were measured with a qubit  fabricated inside a transmission line resonator (circuit QED)~\cite{Wallraff2005}.  Recently, the same design was used to non-destructively measure the photon number population of a resonator~\cite{Schuster2006}.  In flux qubits dephasing times as long as 4 $\mu$s \cite{Bertet2005} were observed. 

A new latching readout based on the non-linear dynamics of a large Josephson junction was implemented~\cite{Siddiqi2006}.  Distinguishing features of this approach are that it is dispersive and that, once the readout as occurred, information about the logical state of the qubit is encoded in the dynamical state of the junction for a long time.  This allows, in principle, to accumulate enough signal about the state of the qubit to overwhelm any noise in the circuitry.  This type of latching readout was also applied to flux qubits~\cite{Lupascu2006}.

{\em Exotic superconducting qubits.---} {\em ``Phase slip" qubit} proposed in Ref.~{\cite{Mooij2005a} replaces the tunneling Josephson junction of an rf SQUID qubit by a thin superconducting wire (bridge), which plays the role of the tunnel barrier.  The geometric inductance of the loop can be kept small, because now the main role is played by the kinetic inductance $L_k$ which relates the supercurrent increase to the change in the superconducting phase difference, $L_k I_s = \frac{\Phi_0 \phi}{2\pi}$. 
{\em ``Andreev level" qubit}~\cite{Zazunov2003}  replaces the tunneling Josephson junction by another type of weak link,  a quantum point contact, i.e., a constriction, in which only a few conducting modes can fit (see ~\cite{Zagoskin1998}, Ch.4). The Andreev levels are formed inside the constriction and carry the Josephson current. 
In either case, the effective tunneling matrix element should not exponentially depend on the microscopic properties of the wire (point contact). It is therefore expected that such qubits would be easier to fabricate, and they would be less sensitive to $1/f$ noise, than a flux qubit.
{\em ``d-wave qubits"} using high-T$_c$ superconductors were proposed in~\cite{Ioffe1999,Blais2000,Amin2005a}. Due to their $d$-wave pairing symmetry, the Josephson junctions with such superconductors may have an {\em intrinsically degenerate} ground state and therefore do not require external circuitry to stay at the optimal point. The ground state degeneracy in YBCO Josephson junctions was
 experimentally verified in Ref.~\cite{Ilichev2001}.  The feared decoherence from nodal quasiparticles could have been  overestimated~\cite{Zagoskin2003}. Recent observations of macroscopic quantum tunneling in $d$-wave junctions based on YBCO \cite{Bauch2005} and BiSCCO \cite{Inomata2005,Jin2006}, as well as coherent transitions in an YBCO junction \cite{Bauch2006}, lend credibility to these conclusions and make the possibility of high-T$_c$ qubits more realistic. Interestingly, YBCO junctions with $\pi$-phase shifts were also recently incorporated in RSFQ devices~\cite{Ortlepp2006}, which, of course, work in a classical regime.
 Due to the interplay of magnetic and superconducting ordering, a Josephson junction with a {\em ferromagnetic barrier} can have an equilibrium phase difference $\pi$ \cite{Blatter2001,Yamashita2005}. Inserting such a junction in a flux qubit loop (a {\em ``SFS qubit"}) is equivalent to putting in it a $\Phi_0/2$-flux quantum and thus produces the intrinsic degeneracy of the ground. 
 
In conclusion, we can confidently state that superconducting qubits and more general SQCs have a serious potential as ``quantum hardware"; but  it is their unique suitability for probing the boundaries between quantum and classical behaviour of nature, which makes them  so attractive objects of research.

\bibliography{REVIEW_0108bisFINresp2c}
 
\bibliographystyle{apsrev}

\begin{figure}[b]
\includegraphics[width=.4\textwidth]{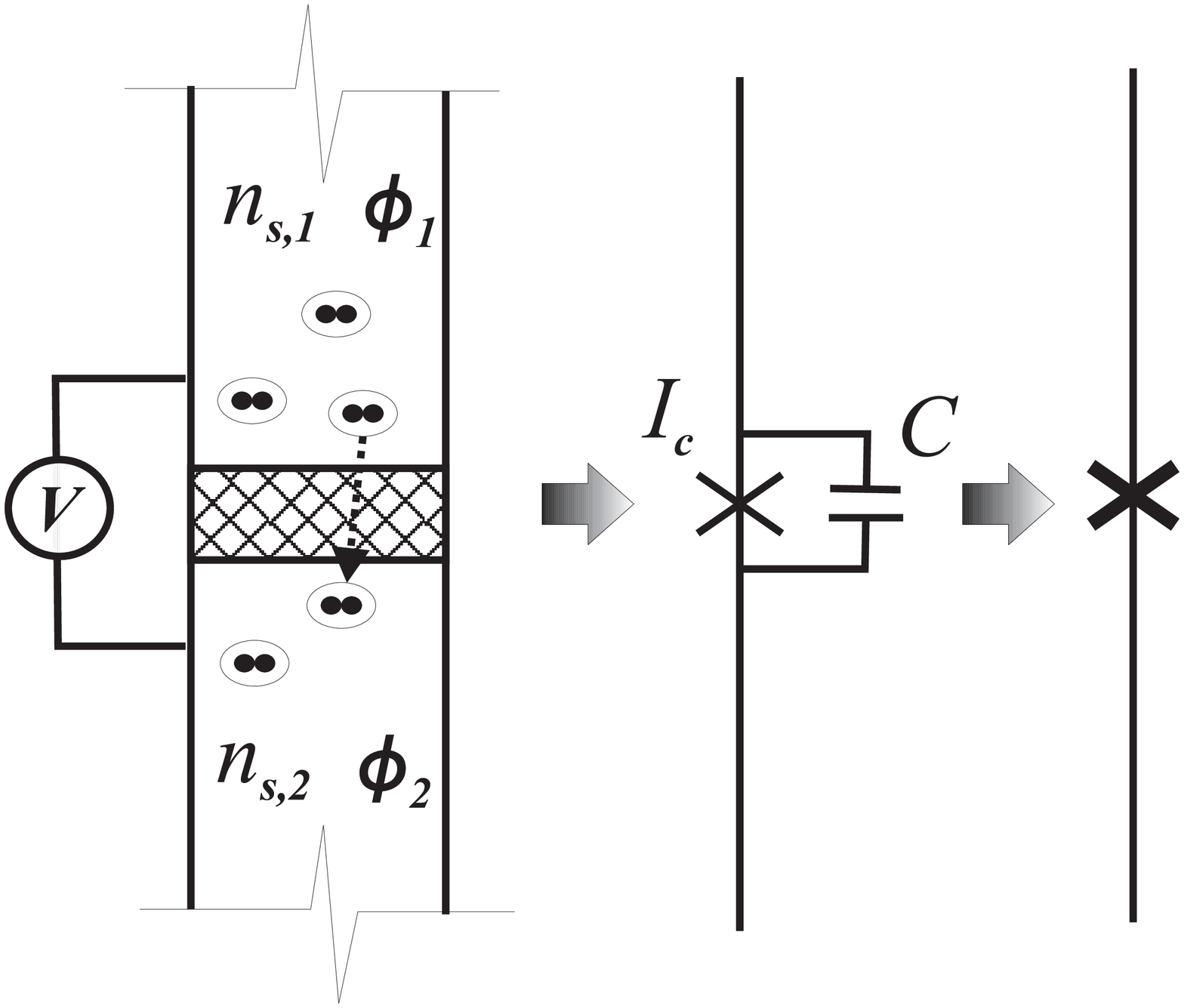}
\caption{Josephson junction and Josephson effect. The state of the $j$th superconductor is characterized by the ``superconducting electron density", $n_{s,j}$, and superconducting phase, $\phi_j$. In case of tunneling Josephson junction, shown here, both the critical current $I_c$ of the junction and its effective  capacitance $C$ depend on thickness of the insulating layer, through which the Cooper pairs tunnel.}
\label{fig_jj0}
\end{figure}

\begin{figure}
\includegraphics[width=.4\textwidth]{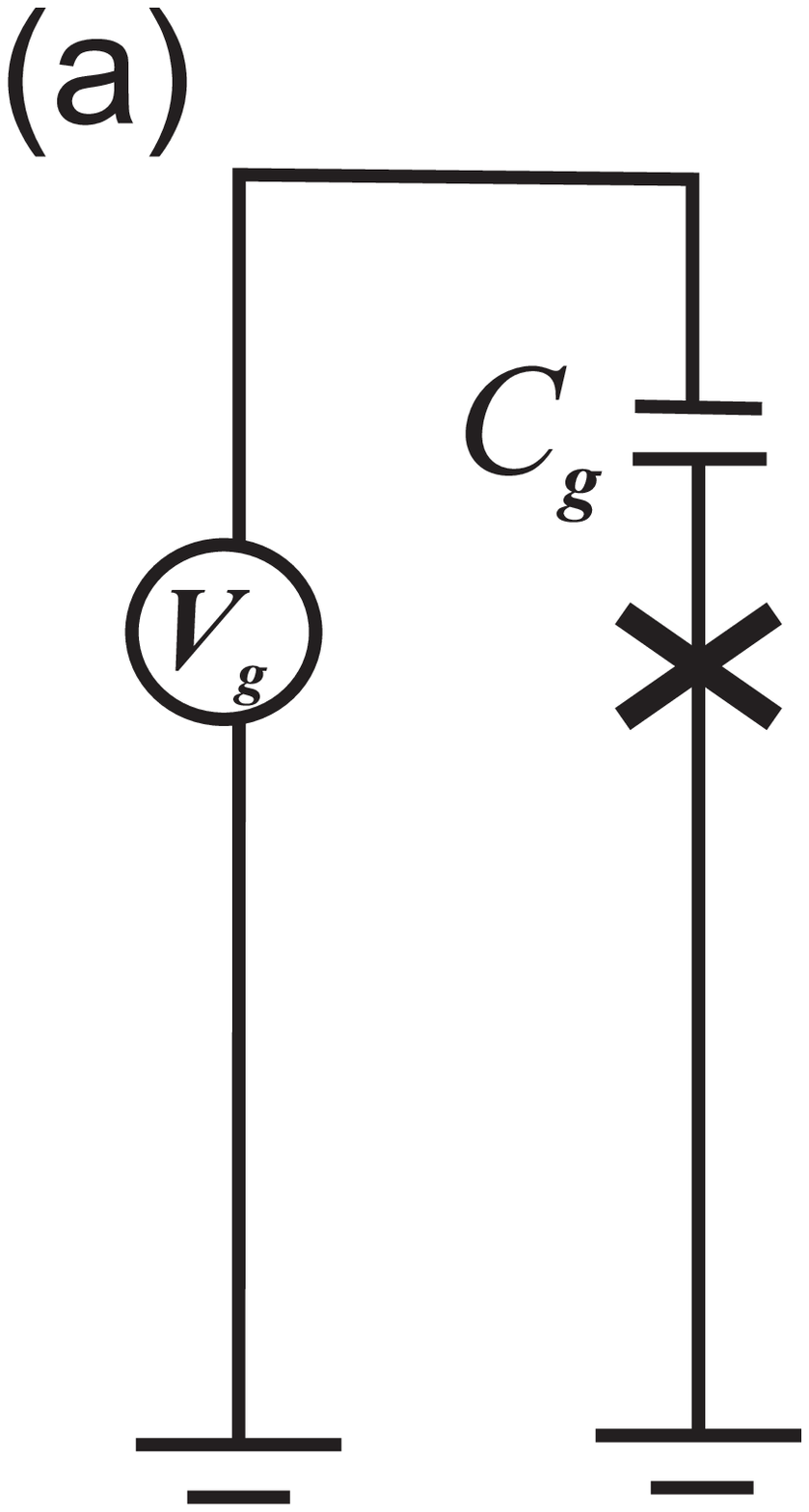}
\includegraphics[width=.4\textwidth]{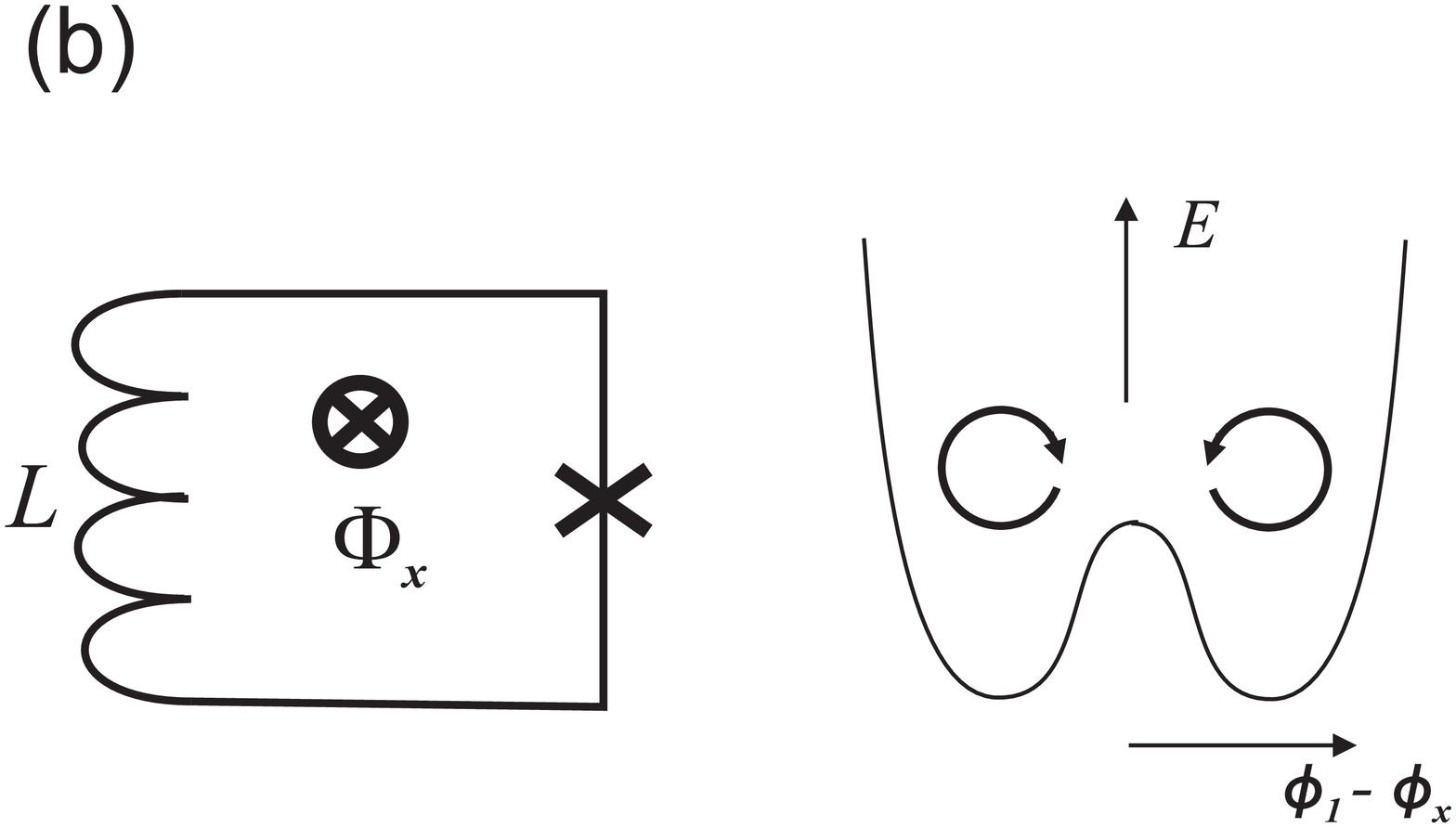}
\includegraphics[width=.4\textwidth]{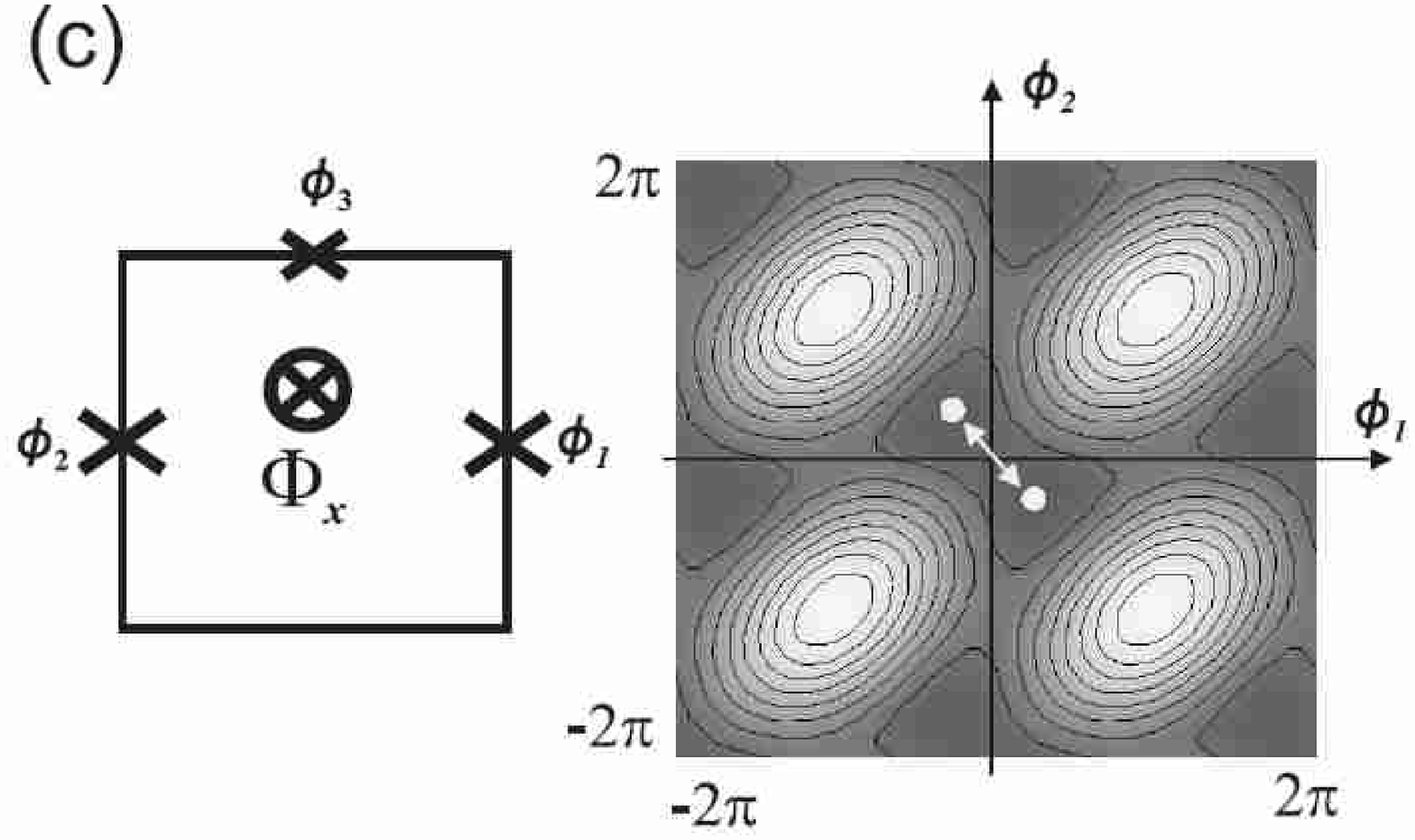}
\includegraphics[width=.4\textwidth]{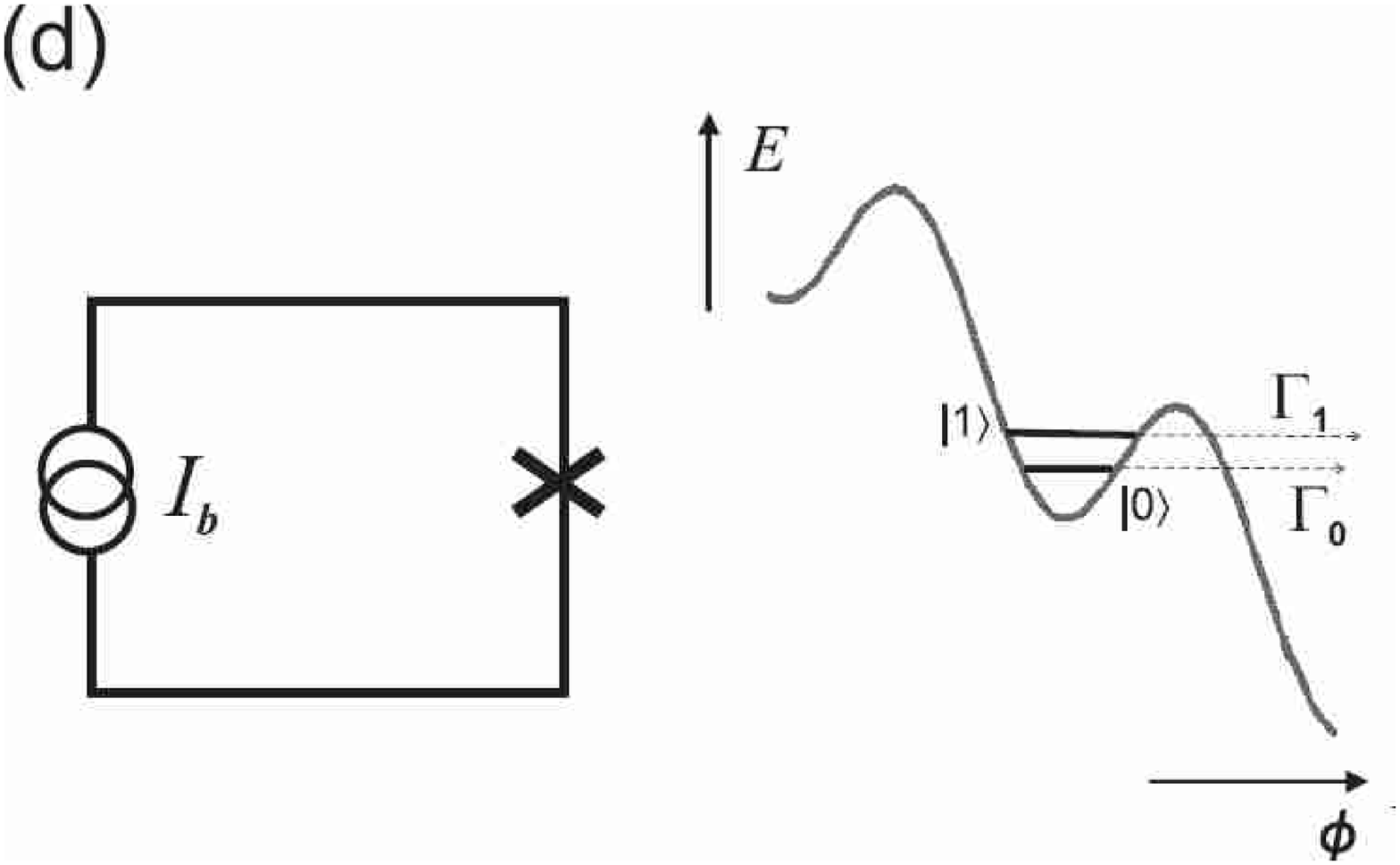}
\caption{(a) Simplest charge qubit \cite{Nakamura1999}. (b) RF SQUID (single-junction flux qubit \cite{Friedman2000}) and its potential energy, when the external magnetic flux through the  loop $\Phi_x = \Phi_0/2$. The flux quantum $\Phi_0 = h/2e$.  (c) Three-junction flux qubit \cite{Wal2000a} and its Josephson potential profile as a function of two independent phase differences (see text) when $\Phi_x = \Phi_0/2$. Dots mark degenerate states, between which tunneling occurs. (d) Phase qubit \cite{Martinis2002,Yu2002} and its Josephson energy (``washboard potential"). Two metastable states indicated can be used as qubit states $|0\rangle$ and $|1\rangle$; $\Gamma_0$ and $\Gamma_1 \gg \Gamma_0$ are the tunneling rates out of these states into the continuum.}
\label{fig_jj}
\end{figure}

\begin{figure}
\includegraphics[width=.45\textwidth]{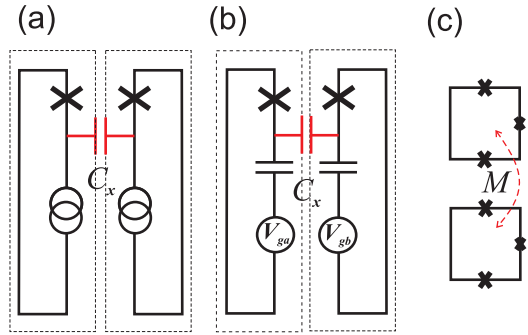}
\caption{(a) Phase qubits and (b) charge qubits coupled through a capacitor. (c) Flux qubits coupled through the mutual inductance $M$.}
\label{fig_coupled_passive}
\end{figure}

\begin{figure}
\includegraphics[width=.45\textwidth]{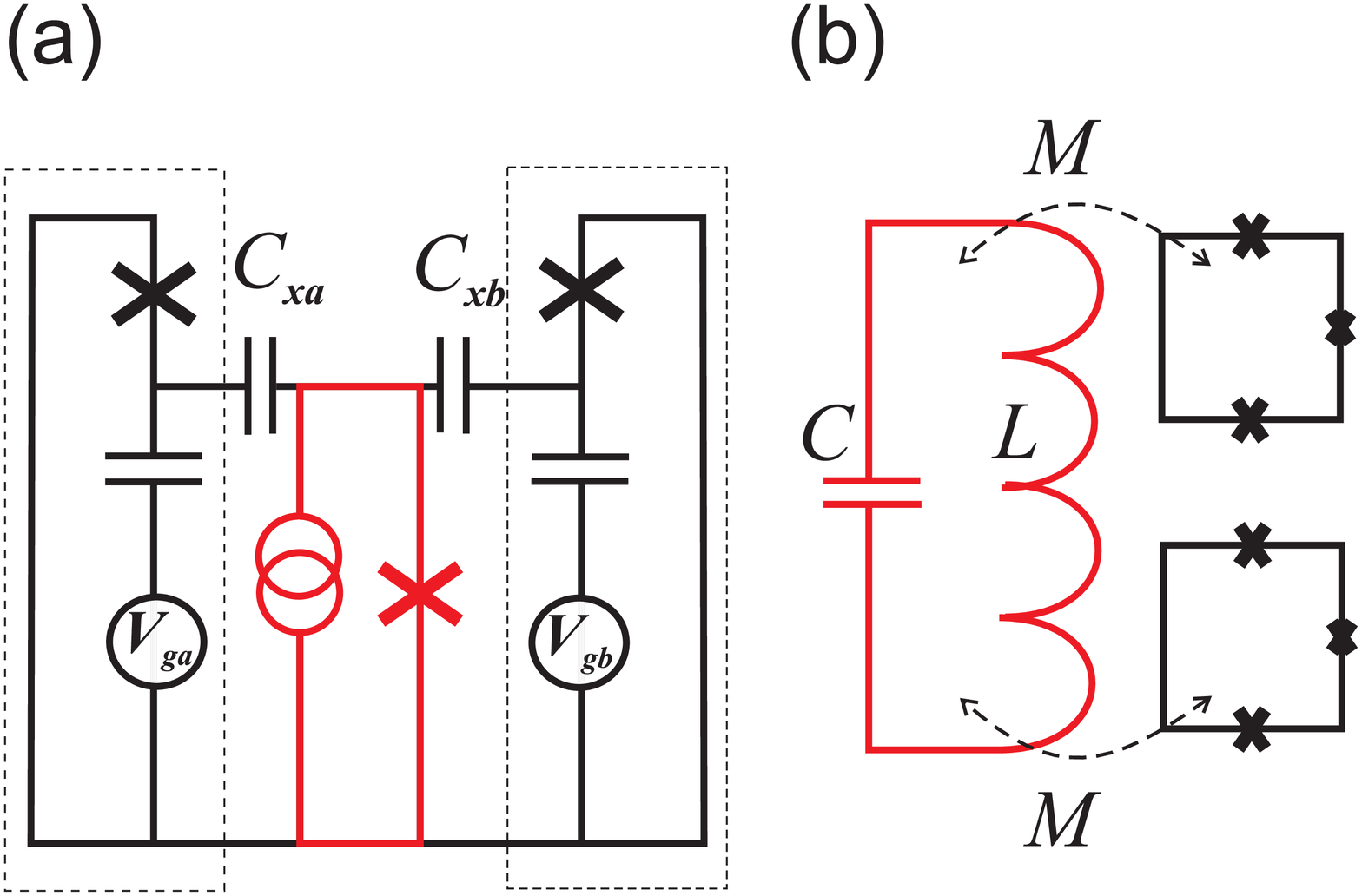}
\includegraphics[width=.45\textwidth]{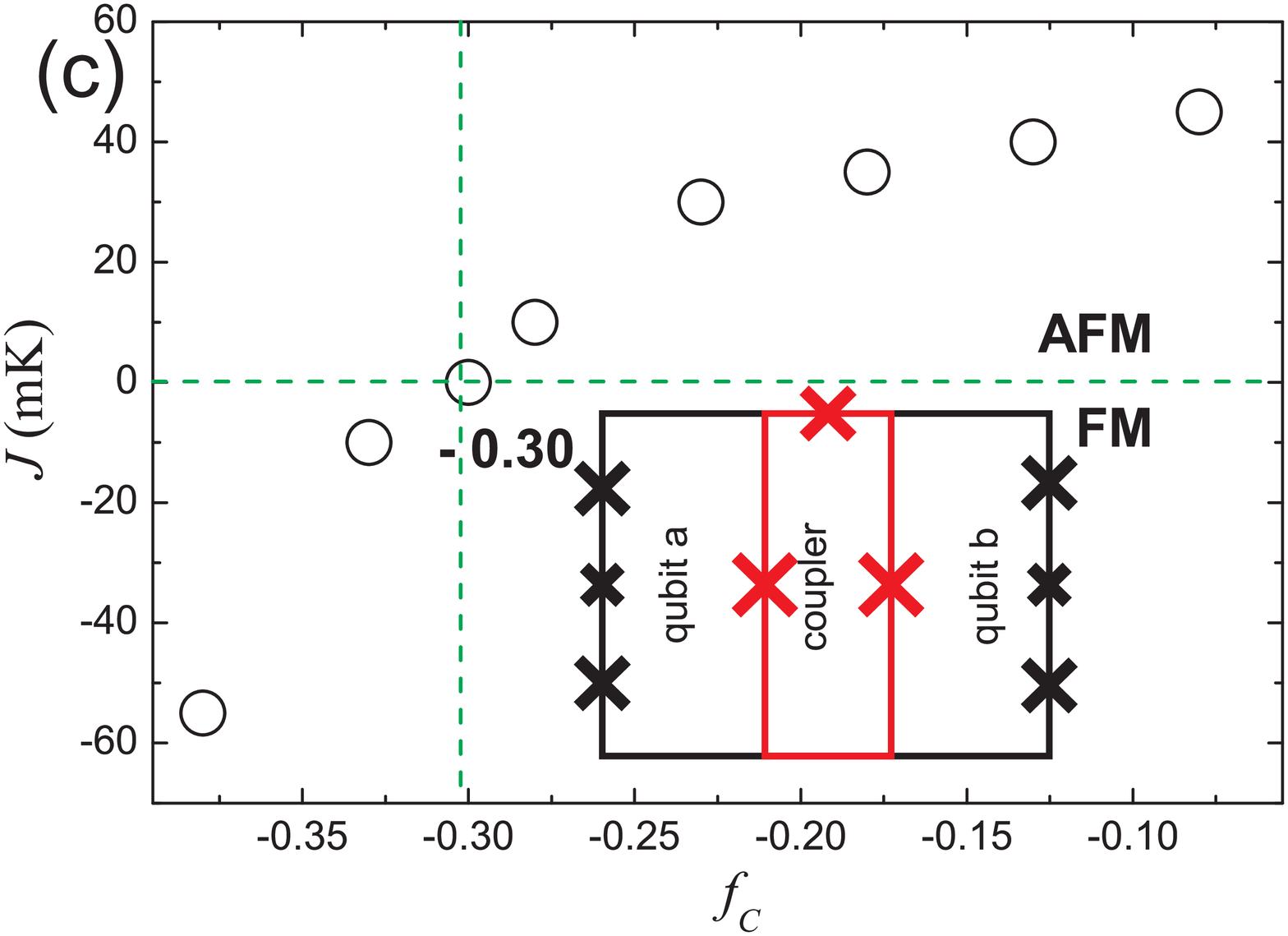}
\caption{(a) Charge qubits coupled through a tunable bus circuit (red) (adapted from Ref.~\cite{Blais2003}). (b) Flux qubits coupled through an $LC$ circuit (red). (c) Tunable coupling of two flux qubits (adapted from Ref.~\cite{Ploeg2006}). The coupling is tuned between ferro- and antiferromagnetic by changing the magnetic flux, $\Phi = f_C \Phi_0$, through the coupler (red). Dots show the experimental data.}
\label{fig_coupled_active}
\end{figure}

\begin{figure}
\includegraphics[width=.45\textwidth]{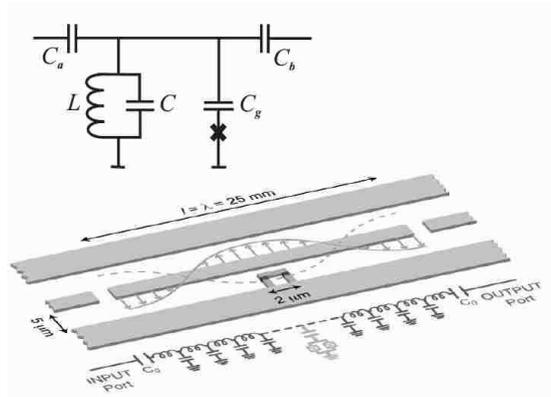}
\caption{Circuit QED (adapted from Ref.\cite{blais:2004}): a charge qubit coupled to a strip line, and the simplified scheme of the system (inset).}
\label{fig_circuit_QED}
\end{figure} 

\end{document}